\documentclass[10pt,letterpaper]{article}
\usepackage{opex3}

\begin{document}

\title{ Dual-wavelength active optical clock}

\author{Zhichao Xu, Duo Pan, Wei Zhuang, and Jingbiao Chen$^*$}

\address{State Key Laboratory of Advanced Optical Communication Systems and Networks, \\Institute of Quantum Electronics, School of Electronics Engineering $\&$ Computer Science, \\Peking University, Beijing 100871, China}

\email{$^*$jbchen@pku.edu.cn}


\begin{abstract}
We experimentally realize the dual-wavelength active optical clock for the first time. As the Cs cell temperature is kept between 118 $^{\circ }C$ and 144 $^{\circ }C$, both the 1359 nm and the 1470 nm stimulated emission output of Cs four-level active optical clock are detected. The 1470 nm output linewidth of each experimental setup of Cs four-level active optical clock is measured to be 590 Hz with the main cavity length unstabilized. To stabilize the cavity length of active optical clock, the experimental scheme of 633 nm and 1359 nm good-bad cavity dual-wavelength active optical clock is proposed, where 633 nm and 1359 nm stimulated emission is working at good-cavity and bad-cavity regime respectively. The cavity length is stabilized by locking the 633 nm output frequency to a super-cavity with the Pound-Drever-Hall (PDH) technique. The frequency stability of 1359 nm bad-cavity stimulated emission output is then expected to be further improved by at least 1 order of magnitude than the 633 nm PDH system due to the suppressed cavity pulling effect of active optical clock, and the quantum limited linewidth of 1359 nm output is estimated to be 77.6 mHz.
\end{abstract}




\section{Introduction}
Recently great progress has been made in single ion optical clocks~\cite{Chou, Huntemann, Pierre, Gao, Margolis} and optical lattice clocks~\cite{Takamoto,Nicholson,Middelmann,McFerran, Hinkly, Bloom, Katori1}. The $10^{-18}$ level accuracy and stability has been experimentally realized. However, all these optical clocks are working in the passive regime. A probing laser prestabilized to a super-cavity with the Pound-Drever-Hall (PDH) technique~\cite{Drever, Young, Holger, Jiang, Kessler} is utilized as the local oscillator in these optical clocks. It is still a great challenge to realize millihertz level linewidth~\cite{Yu1} in optical clock due to the cavity-length noise induced by the thermal Brownian-motion.

Unlike the traditional passive optical clocks, active optical clocks~\cite{Kuppens-1,Chen,
Zhuang, Zhuang1, Zhuang2, Yu2, Chen1, Chen2, Wang, Meiser, Sterr,
Meiser1, Meiser2, Yu3, Yu4, Xie, Zhuang3, Zhuang4, Zhuang5, Li,
Bohnet, Bohnet1, Xue, Zang, Kazakov, Bohnet2,Bohnet3,Zhang, Shengnan,
Yanfei, Xu, Xu2, Pan, Yanfei2, Yanfei3, Dongying} utilize the stimulated emission output of the unperturbed clock transition as optical frequency standards, and the super-cavity stabilized probing lasers are thus unnecessary. Population inversion is established between the two levels of the clock transition and a cavity (the main cavity) is employed to provide weak optical feedback to the stimulated emission of active optical clock. The linewidth of the main cavity mode of active optical clock is designed to be much wider than the gain profile of clock transition, and the frequency of the stimulated emission will thus be insensitive to the main cavity-length noise.

Currently the 1470 nm lasing of Cs four-level active optical clock has been realized~\cite{Xu, Xu2, Pan}. The cavity pulling effect is measured to be reduced by a factor of about 40~\cite{Xu, Pan} and a linewidth of sub-kHz is observed. However, the main cavity length of Cs active optical clock is still not stabilized.

In this paper, the dual-wavelength active optical clock is experimentally realized for the first time. When the Cs cell temperature is kept between 118 $^{\circ }C$ and 144 $^{\circ }C$, both the 1470 nm and the 1359 nm stimulated emission output of Cs four-level active optical clock are detected. Two independent and uncorrelated experimental setups of Cs four-level active optial clock which are respectively pumped by 455 nm and 459 nm external-cavity diode laser (ECDL) are established. The 1470 nm output linewidth of each experimental setup of Cs four-level active optical clock is measured to be 590 Hz by beating and comparing these two independent setups. However, the main cavity length of Cs four-level active optical clock is not stabilized and the frequency stability of 1470 nm stimulated emission output is greatly affected by the perturbation from external environment. In order to stabilize the main cavity length, the experimental scheme of 633 nm and 1359 nm good-bad cavity dual-wavelength active optical clock is proposed. The 633 nm stimulated emission is designed to work in the good-cavity limit, i.e. the 633 nm cavity mode linewidth is designed to be much narrower than the 1.5 GHz 633 nm gain bandwidth, and is frequency stabilized to a super-cavity with PDH technique. The main cavity of the good-bad cavity active optical clock is then stabilized, and the stability of 1359 nm bad-cavity stimulated emission output frequency of active optical clock is expected to be improved by at least 1 order of magnitude than that of the 633 nm PDH stabilized stimulated emission output. The quantum limited linewidth of 1359 nm stimulated emission output of the good-bad cavity active optical clock is estimated to be 77.6 mHz.

\section{Experiment}
The four-level configuration of Cs atoms is employed in the dual-wavelength active optical clock and the relevant energy levels are shown in Fig.1. Cs atoms of the ground state 6S$_{1/2}$ are excited to the 7P$_{3/2}$ state by a 455 nm pumping laser. The Cs atoms pumped to the 7P$_{3/2}$ state will then decay to the lower states such as 6S$_{1/2}$, 7S$_{1/2}$, 5D$_{3/2}$ and 5D$_{5/2}$. In the steady state, 5.8\% of Cs atoms are in the 7S$_{1/2}$ state, while 1.4\% and 2.9\% of Cs atoms are in the state of 6P$_{1/2}$ and 6P$_{3/2}$ respectively~\cite{Yanfei}. The population inversion between 7S$_{1/2}$ and 6P$_{1/2}$ for 1359 nm clock transition7S$_{1/2}$, and 7S$_{1/2}$ and 6P$_{3/2}$ for 1470 nm clock transition, is thus established. When the population inverted Cs atoms are placed in a specially designed optical cavity with the cavity mode linewidth much wider than the gain bandwidth of Cs atoms, stimulated emission output at 1359 nm and 1470 nm will be observed once the stimulated emission threshold is met. The stimulated emission of 1359 nm and 1470 nm with the weak optical feedback from the cavity is thus the active optical clock output. It is notable that the 1360 nm transition between 7P$_{3/2}$ and 5D$_{5/2}$ states is close in wavelength to the 1359 nm transition between 7S$_{1/2}$ and 6P$_{1/2}$ states. However, in the steady state 25.4\% of Cs atoms are in the 7P$_{3/2}$ state, while 35.8\% of Cs atoms are in the 5D$_{5/2}$ state~\cite{Yanfei}. The population inversion is not realized between 7P$_{3/2}$ and 5D$_{5/2}$ states because the lifetime of 5D$_{5/2}$ is longer than that of  7P$_{3/2}$ , and the corresponding 1360 nm stimulated emission can not be detected.
\begin{figure}[htbp]
\centering\includegraphics[width=12cm]{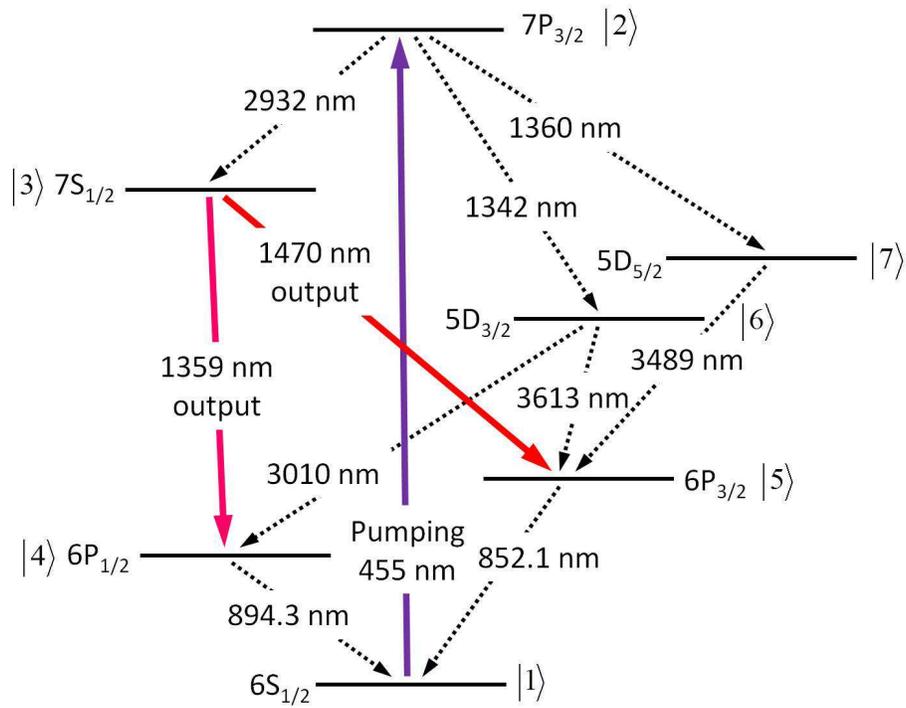}\\
\caption{(color online). Related energy levels of Cs. For simplicity, each level is labeled as~$ |i\rangle $.}\label{Figure1}
\end{figure}

The experimental setup of the dual-wavelength Cs four-level active optical clock is illustrated in Fig.2.  A 455 nm external-cavity diode laser (ECDL) frequency stabilized to Cs 455 nm saturated spectrum is used as the pumping laser. M3 and M4 shown in Fig.2 are 455 nm high-reflecting mirrors used to introduce the pumping laser beam into the main cavity of the dual-wavelength Cs four-level active optical clock, which is composed of a plane mirror (M1 in Fig.2) and a concave mirror (M2 in Fig.2). The main cavity length is set to be 8.6 cm, and the free spectrum range (FSR) is then 1.74 GHz. M2 is coated with 455 nm anti-reflection and 1359 nm/1470 nm high-reflection coating, and the radius of curvature of M2 is set to be 8000 mm. M1 is coated 455 nm anti-reflection coating and the reflectivity at 1359 nm/1470 nm is set to be 80\%. The transmission of the Cs cell placed in the main cavity at 1359 nm/1470 nm is measured to be about 80\%. The main cavity length of the dual-wavelength Cs four-level active optical clock is controlled by the piezoelectric ceramic transducer (PZT) installed on M1. The output power of the dual-wavelength Cs four-level active optical clock is measured by the Thorlabs PDA50B-EC photodetector at 800-1800 nm (PD shown in Fig.2).
\begin{figure}[htbp]
\centering\includegraphics[width=12cm]{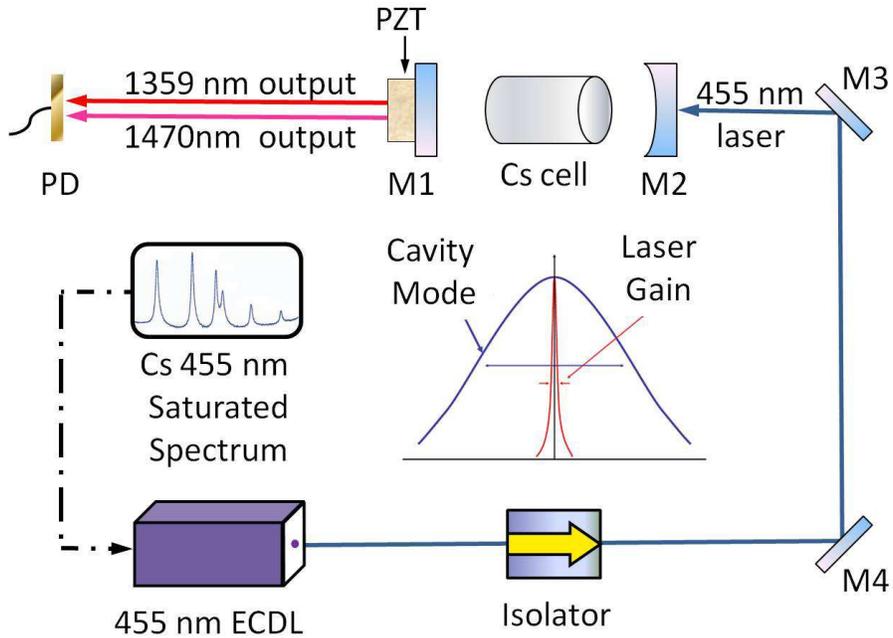}\\
\caption{(color online). Experimental setup of the dual-wavelength Cs four-level active optical clock.}\label{Figure2}
\end{figure}

According to our previous experimental results, the maximum output power of the dual-wavelength Cs four-level active optical clock is obtained when the 455 nm pumping laser frequency is locked to the Cs 455 nm 6S$_{1/2}$ (F=4) and 7P$_{3/2}$ (F'=5) transition~\cite{Xu2}. The 455 nm pumping laser power introduced into the main cavity is measured to be 11.77 mW. The velocity selective pumping scheme where the 455 nm pumping laser beam is aligned parallel to the 1359 nm/1470 nm main cavity mode is employed to reduce the laser gain bandwidth to about 9 MHz, while the linewidth of the main cavity mode is measured to be more than 400 MHz~\cite{Xu2}. The laser gain bandwidth is then much narrower than the main cavity mode linewidth as shown in Fig.2, and the dual-wavelength Cs four-level active optical clock is thus working in the bad-cavtiy regime. The cavity pulling effect of the dual-wavelength Cs four-level active optical clock is reduced by a factor of about 40~\cite{Xu2} and the output frequency then mainly depends on the Cs clock transition instead of the unstable macroscopic main cavity length.

The Thorlabs FEL1450 filter is used to eliminate the 1359 nm output signal before the 1470 nm output power of the dual-wavelength Cs four-level active optical clock is measured by the Thorlabs PDA50B-EC photodetector (represented by PD in Fig.2). The transmission of the Thorlabs FEL1450 filter at 1470 nm is measured to be 90\% and no transmitted 1359 nm signal is detected. The main cavity length of the dual-wavelength Cs four-level active optical clock is kept in resonance with the 1470 nm output to maximize the 1470 nm output power as the 455 nm pumping laser power and the Cs cell temperature is fixed. The Cs cell temperature is changed from 80 $^{\circ }C$ to 170 $^{\circ }C$ and the 1470 nm output power is measured. Then the Thorlabs FEL1450 filter is removed and the 1359 nm \& 1470 nm total output power of the dual-wavelength Cs four-level active optical clock can be  measured.

\section{Experimental results}
\begin{figure}[htbp]
\centering\includegraphics[width=12cm]{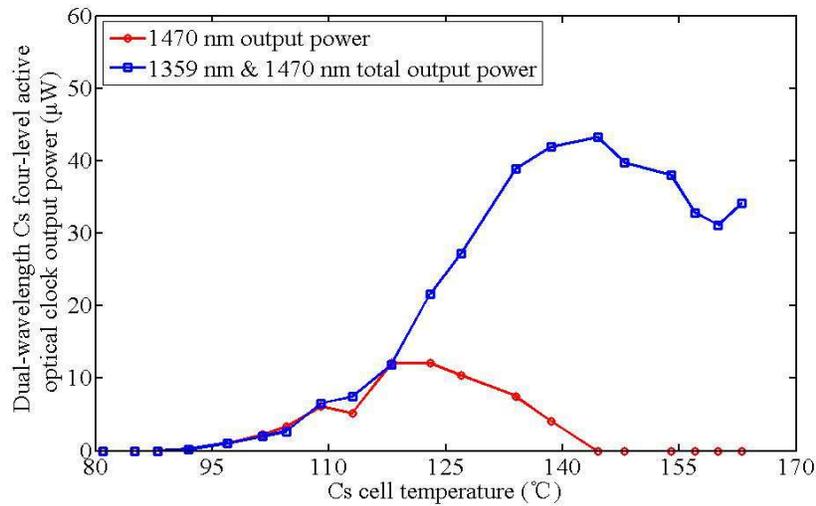}\\
\caption{(color online). Measured 1470 nm (represented by the red curve) and the total output power (represented by the blue curve) of the dual-wavelength Cs four-level active optical clock when changing the Cs cell temperature.}\label{Figure3}
\end{figure}
The measured 1470 nm output power of the dual-wavelength Cs four-level active optical clock when changing the Cs cell temperature is represented by the red curve in Fig.3, while the total output power, i.e. the sum of 1359 nm and 1470 nm output power, is represented by the blue curve in Fig.3. As mentioned above, the 455 nm pumping laser frequency is locked to the Cs 455 nm 6S$_{1/2}$ (F=4) and 7P$_{3/2}$ (F'=5) transition~\cite{Xu2} and the pumping laser power is set to be 11.77 mW. When the Cs cell temperature is lower than 92 $^{\circ }C$, the laser threshold is not satisfied and no stimulated emission output of the dual-wavelength Cs four-level active optical clock can be detected. As the Cs cell temperature is increased from 92 $^{\circ }C$ to 118 $^{\circ }C$, the 1470 nm stimulated emission output is found to gradually increase from 0 $\mu$W to about 12 $\mu$W, while the 1359 nm stimulated emission output is still not detected. In this case the measured total output power of the dual-wavelength Cs four-level active optical clock is thus found to equal to the 1470 nm output power, as shown in Fig.3. As the Cs cell temperature is kept between 118 $^{\circ }C$ and 144 $^{\circ }C$, both the 1470 nm stimulated emission output and the 1359 nm stimulated emission output will be detected, and the measured total output power will be larger than the 1470 nm output power. The Cs four-level active optical clock is then found to work in the dual-wavelength output regime. This is the first time that the dual-wavelength active optical clock is experimentally realized. The 1470 nm output power has a maximum value of approximately 12.1 $\mu$W near the 120 $^{\circ }C$ Cs cell temperature and will then gradually decrease to 0 $\mu$W at 144 $^{\circ }C$. In contrast, the 1359 nm output power is measured to increase from 0 $\mu$W at 118 $^{\circ }C$ to the maximum 43.2 $\mu$W at 144 $^{\circ }C$. The mode competition between 1470 nm and 1359 nm stimulated emission then appears, and the increase of 1359 nm output power results in the reduction of 1470 nm output power. This is because the 1359 nm stimulated emission and the 1470 nm stimulated emission share the same laser gain medium, i.e. the Cs atoms excited to the 7S$_{1/2}$ state in the heated glass cell. When the Cs cell temperature is higher than 144 $^{\circ }C$, the 1470 nm stimulated emission output will vanish and the 1359 nm output power will decrease with the Cs cell temperature.
\begin{figure}[htbp]
\centering\includegraphics[width=12cm]{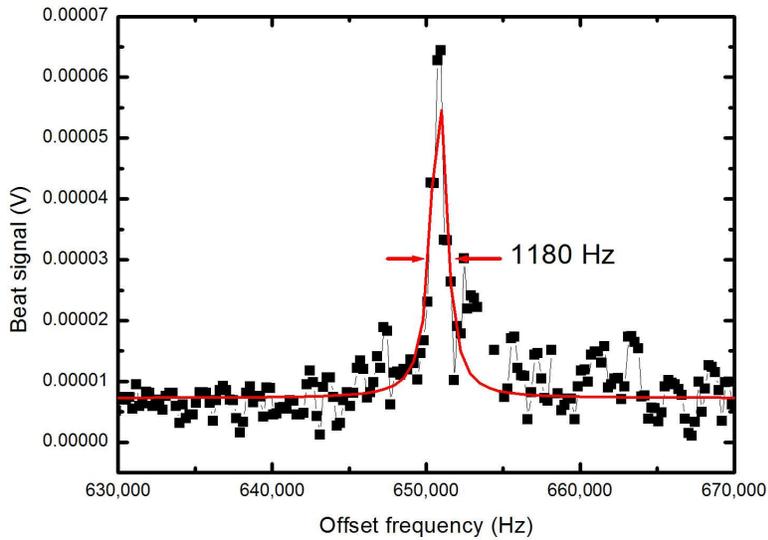}\\
\caption{(color online). 1470 nm beat signal between two independent setups of Cs four-level active optical clock respectively pumped by 455 nm and 459 nm laser. The 1470 nm stimulated emission output linewidth of each experimental setup of Cs four-level active optical clock is 590 Hz.}\label{Figure4}
\end{figure}

To eliminate the influence of magnetic field, magnetic shield is employed and Zeeman splitting is thus not considered in this paper. The hyperfine states F=3, 4 of 7S$_{1/2}$ and F'=3, 4 of 6P$_{1/2}$ can be involved in the newly observed 1359 nm stimulated emission output. The emission frequency difference between F=3~F'=4 transition and F=4-F'=3 transition is 3.35 GHz, approximately 2 times of the 1.74 GHz FSR of the main cavity. Both of the transitions can thus be involved in the 1359 nm stimulated emission output by tuning the main cavity length. Further research is still needed to determine which hyperfine states are involved in the 1359 nm stimulated emission output.

To evaluate the 1470 nm stimulated emission output linewidth, the second setup of Cs four-level active optical clock as shown in Fig.2 is established. The 455 nm pumping laser shown in Fig.2 is replaced by a 459 nm ECDL in the second experimenal setup. The 1470 nm stimulated emission output linewidth is measured by beating and comparing these two equal, independent and uncorelated systems, and the beat signal from the spectrum analyzer is fitted with the Lorentzian profile (represented by the red line) as shown in Fig.4. The swept frequency range is set to be 100 kHz within 254 ms while the RBW is 300 Hz. The 1470 nm stimulated emission output linewidth of each experimental setup of Cs four-level active optical clock is 590 Hz, $1/2$ of the 1180 Hz FWHM of the beat signal fitted with the Lorentzian profile shown in Fig.4. However, the main cavity length of the Cs four-level active optical clock is not stabilized and the frequency stability of 1470 nm stimulated emission output is greatly affected by the perturbation from external environment.

The dual-wavelength Cs four-level active optical clock can be utilized to establish the active optical clock system with its main cavity length locked using the Pound-Drever-Hall (PDH) technique. If the main cavity of the dual-wavelength Cs four-level active optical clock is specially designed so that 1359 nm stimulated emission works in the bad-cavity regime while the 1470 nm stimulated emission works in the good-cavity regime, the main cavity length can thus be stabilized by locking the 1470 nm output frequency to a ultra-stable high-finesse cavity with the PDH technique. The stability of 1359 nm output frequency will then be improved by at least 1 order of magnitude than the 1470 nm PDH frequency stabilization system due to the suppressed cavity-pulling effect of bad-cavity 1359 nm output in Cs four-level active optical clock~\cite{Xu2}. However, the mode competition between 1470 nm and 1359 nm stimulated emission is a problem. To avoid the mode competition between the two output wavelengths of the dual-wavelength active optical clock, the experimental scheme of 633 nm and 1359 nm good-bad cavity dual-wavelength active optical clock is proposed in this paper. As shown in Fig.3, the maximum output power of 1359 nm stimulated emission is much higher than that of 1470 nm stimulated emission, therefore the stimulated emission output of 1359 nm instead of that of 1470 nm is chosen to be the optical frequency standard in the good-bad cavity dual-wavelength active optical clock.

\section{Experimental scheme of 633 nm and 1359 nm good-bad cavity dual-wavelength active optical clock}
As mentioned above, the suppressed cavity pulling effect is the uppermost advantage of active optical clock. If the main cavity length of active optical clock is locked to a ultra-stable high-finesse cavity with the Pounder-Drever-Hall (PDH) technique, the output frequency stability of active optical clock is expected to be further improved by at least 1 to 2 orders of magnitude than the PDH system due to the suppressed cavity pulling effect in active optical clock. Therefore, we proposed the experimental scheme of 633 nm and 1359 nm good-bad cavity dual-wavelength active optical clock.

As illustrated in Fig.5, M1 is a plane mirror coated with 633 nm and 1359 nm highly reflection and 455 nm anti-reflection coating. M2 is the Brewster window coated with 633 nm, 1359 nm and 455 nm anti-reflection coating. M3 is a concave mirror coated with 455 nm anti-reflection coating and the radius of curvature is chosen to be Radius=16 m. The reflectivity of M3 at 633 nm and 1359 nm is respectively chosen to be R$_{633}$=96\% and R$_{1359}$=80\%. According to our experimental results, the transmission at 633 nm and 1359 nm of the Cs cell is set to be T$_{633}$=90\% and T$_{1359}$=80\%. Unlike the main cavity of the dual-wavelength Cs four-level active optical clock shown in Fig.2, an extra specially designed He-Ne laser element is installed within the main cavity of the 633 nm and 1359 nm good-bad cavity dual-wavelength active optical clock. The He-Ne gas mixture and the Cs vapor in the heated cell in Fig.5 are respectively used as 633 nm and 1359 nm laser medium. As a result, the increase of the 633 nm output power will not influence the output power of 1359 nm active optical clock output for their gain medium are chosen to be different elements, and the mode competition of the two output wavelength shown in Fig.3 will then be avoided. The main cavity of the 633 nm and 1359 nm good-bad cavity dual-wavelength active optical clock is then composed of a plane mirror (M1 in Fig.5) and a concave mirror (M3 in Fig.5). A piezoelectric ceramic transducer (PZT) is installed on M3 to control the main cavity length. The 633 nm and 1359 nm output of the good-bad cavity dual-wavelength active optical clock will share the same main cavity and their cavity modes will spatially overlap each other in the main cavity.

The population inversion of 633 nm output is realized by electrical pumping as in the traditional He-Ne laser system, while a 455 nm external-cavity diode laser (ECDL) is chosen to be the pumping laser for 1359 nm stimulated emission. The frequency of 455 nm pumping laser is locked to the transition of 6S$_{1/2}$ (F=4) and 7P$_{3/2}$ (F'=5) using the Cs 455 nm saturated absorption spectrum. M4 and M5 are 455 nm high-reflecting mirrors and are used to introduce the 455 nm pumping laser into the main cavity of the 633 nm and 1359 nm good-bad cavity dual-wavelength active optical clock. M6 is coated with 455 nm anti-reflection and 633 nm $\&$ 1359 nm high-reflection coating, while M7 is coated with 633 nm anti-reflection and 1359 nm high-reflection coating. PD1 is a photodetector at 1359 nm and PD2 is a photodetector at 633 nm. The 633 nm output is reflected by a high-reflecting mirror at 633 nm (M8) and is used for frequency stabilization with PDH technique. The EOM in Fig.4 is the electro-optical modulator and the ULE cavity is the super-cavity made from ultralow expansion (ULE) glass.
\begin{figure}[htbp]
\centering\includegraphics[width=12cm]{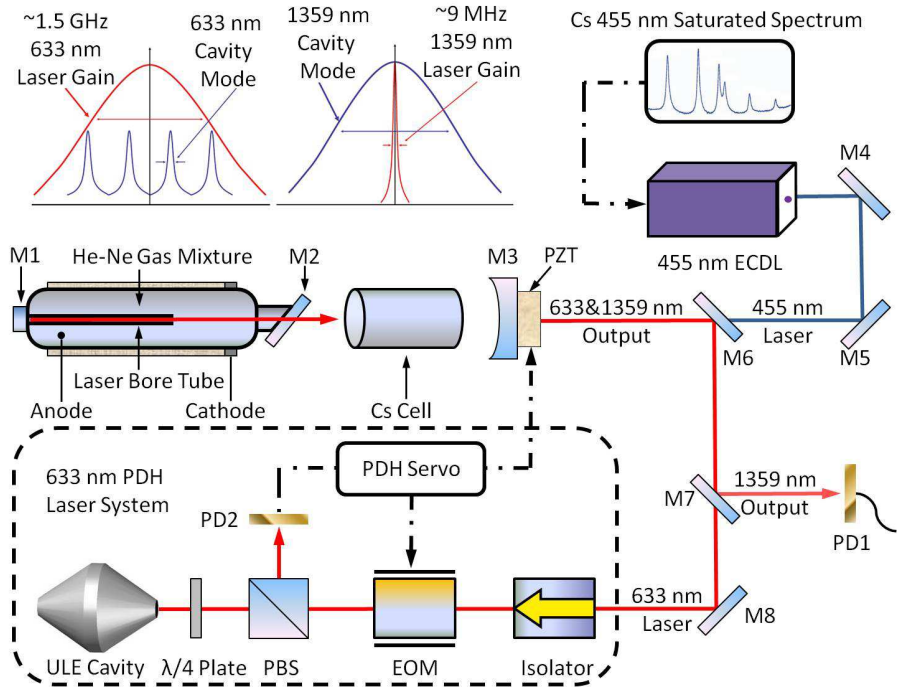}\\
\caption{(color online). Experimental scheme of 633 nm and 1359 nm good-bad cavity dual-wavelength active optical clock.}\label{Figure5}
\end{figure}

The 633 nm output of the good-bad cavity dual-wavelength active optical clock is designed to work in the good-cavity regime, i.e. the 633 nm laser gain bandwidth $\Gamma_{g633}$ is chosen to be much wider than the 633 nm main cavity mode linewidth $\Gamma_{c633}$. Therefore, the 633 nm output frequency will exactly follows the variation of the main cavity length of the good-bad cavity dual-wavelength active optical clock, just as that in traditional lasers. The 633 nm output frequency is stabilized to a ultra-stable high-finesse cavity (the ULE cavity shown in Fig.5) and then the main cavity length of the 633 nm and 1359 nm good-bad cavity dual-wavelength active optical clock will be also exactly locked to the super-cavity. The 1359 nm output of the dual-wavelength active optical clock shares the same main cavity with the 633 nm output and is designed to work in the bad-cavity regime. The 1359 nm main cavity mode linewidth $\Gamma_{c1359}$ is chosen to be approximately 24 times of the 1359 nm laser gain bandwidth $\Gamma_{g1359}$, and the cavity pulling effect of 1359 nm output is expected to be reduced by a factor of about 25~\cite{Xu2}. The 1359 nm output frequency stability of the dual-wavelength active optical clock is then expected to be further improved by at least 1 order of magnitude than the 633 nm PDH system

The main cavity length of the 633 nm and 1359 nm good-bad cavity dual-wavelength active optical clock is designed to be $L=15 $ cm. The length of the Cs cell in the main cavity is $L_{cell}=5 $ cm while the length of the specially designed He-Ne laser element is about 10 cm. The free spectral range (FSR) of the main cavity is $FSR=\frac{c}{2L}=1 $ GHz. The 633 nm gain bandwidth is chosen to be 1.5 GHz, the same as in traditional 633 nm He-Ne lasers. The reflectivity of the main cavity output mirror (M3 in Fig.5) at 633 nm is $R_{633}=96\%$ and the finesse of the main cavity at 633 nm is then calculated to be $F_{633}=\frac{\pi\sqrt{R_{633}{T_{633}}^2}}{1-R_{633}{T_{633}}^2}=12.5$. The 633 nm main cavity mode linewidth is thus $\Gamma_{c633}=\frac{FSR}{F_{633}}=80.0 $ MHz, which is far less than the 1.5 GHz 633 nm laser gain bandwidth $\Gamma_{g633}$. Therefore, the 633 nm output of the good-bad cavity dual-wavelength active optical clock is working in the good-cavity limit and the 633 nm output frequency follows exactly with the main cavity length variation. The frequency of the main cavity mode can thus be exactly locked to the ULE cavity with the 633 nm PDH frequency stabilization system shown in Fig.5.

As mentioned above, the 455 nm pumping laser power is set to be 11.77 mW, while the waist radius of pumping laser beam at the position of main cavity is measured to be $0.463\times0.855 $ mm by using the Newport's LBP-2-USB laser beam profiler. According to our previous work~\cite{Xu2}, the Lorentzian linewidth of Cs 455 nm transition profile caused by saturation broadening of pumping laser is 23.58 MHz, and a velocity spread of 10.74 m/s parallel to the main cavity mode of Cs atoms can be excited to the 7S$_{1/2}$ state. The Lorentzian Doppler linewidth of Cs 1359 nm transition induced by the saturation broadening of 455 nm pumping laser is then calculated to be $\Gamma_{sat}=7.90 $ MHz. The Lorentzian natural linewidth of 1359 nm transition is 0.99 MHz and the 1359 nm laser gain bandwidth is thus calculated to be $\Gamma_{g1359}=8.89 $ MHz~\cite{Xu2}. The finesse of the main cavity at 1359 nm is theoretically calculated to be $F_{1359}=\pi\frac{\sqrt{R_{1359}{T_{1359}}^2}}{1- R_{1359}{T_{1359}}^2}=4.6$. The 1359 nm linewidth of the main cavity mode is then $\Gamma_{c1359}=\frac{FSR}{F_{1359}}=217.39 $ MHz. The factor $a=\frac{\Gamma_{c1359}}{\Gamma_{g1359}}=24.5$ and the cavity pulling effect of 1359 nm output is expected to be reduced by a factor of 25.5 according to our previous experimental results~\cite{Xu2}. The 1359 nm output working in the bad-cavity regime is used as the final optical frequency standard of the 633 nm and 1359 nm good-bad cavity dual-wavelength active optical clock and the frequency stability of the 1359 nm output can be improved by at least 1 order of magnitude than 633 nm PDH system.

\section{Quantum limited linewidth of 633 nm and 1359 nm good-bad cavity dual-wavelength active optical clock }
An estimate of the quantum limited linewidth of the 1359 nm output is given here. The spot size of the 1359 nm main cavity mode is $w_0={[{\frac{\lambda^2}{\pi^2}}L(Radius-L)]}^{1/4}=0.819 $ mm,where $\lambda$ is the wavelength of 1359 nm output of good-bad cavity dual-wavelength active optical clock. The 1359 nm main cavity mode volume is thus calculated to be $V_c={\frac{1}{4}}{\pi}{{w_0}^2}L=7.90\times{10}^{-8} $ m$^3$.The natural linewidth of cesium 1359 nm transition is $\Gamma_{1359}=0.99 $ MHz, and the electric dipole moment of cesium 1359 nm transition is calculated to be $\mu=\sqrt{\frac{3\epsilon_0h\Gamma_{1359}\lambda^3}{8\pi^2}}=2.35\times10^{-29}$ C$\cdot$m. The atom-cavity coupling constant g is thus calculated to be $g={\frac{2\pi \mu}{h}}\sqrt{\frac{h\nu_{1359}}{2\epsilon_0 V_c}}=2\pi \times 11.4 $ kHz, where $\nu_{1359}$ is the Cs 1359 nm transition frequency. The main cavity loss rate at 1359 nm is $\kappa=2\pi\Gamma_{c1359}=1.37\times10^9$ s$^{-1}$. Acording to our previous work~\cite{Xu2}, the average intensity of 455 nm pumping laser within the Cs cell is estimated to be $I=695.2$ mW/cm$^2$ considering the power losses of the main cavity mirrors and Cs cell windows. The Rabi frequency of 455 nm pumping laser is $\Omega=\sqrt{\frac{3{\lambda_{21}}^{3}\gamma_{21}I}{2\pi hc}}=5.5\times10^7$ s$^{-1}$~\cite{Xue}.

As illustrated in Fig. 3, the 1359 nm stimulated emission output power of Cs four-level active optical clock reaches its maximum value at the 144 $^{\circ }C$ Cs cell temperature, and the 1470 nm stimulated emission output then vanished. Therefore, the Cs cell temperature of 633 nm and 1359 nm good-bad cavity dual-wavelength active optical clock is set to be 144 $^{\circ }C$ here, and the vapor pressure of Cs atoms in the glass cell is $\log P_V=11.0531-\frac{4041}{T}-1.35\log T$~\cite{Nesmeyanov}, where $P_V$ is the Cs vapor pressure in torr, and $T=417$ K is the absolute temperature of the Cs cell. The Cs vapor pressure is then calculated to be $P=0.89$ Pa and the Cs atomic density in the Cs cell is $n=\frac{P}{k_BT}=1.55\times10^{20}$ /m$^3$, where $k_B$ is the Boltzmann constant. The number of Cs atoms in the main cavity mode is $N=\frac{1}{4}n\pi {w_0}^2 L_{cell}=4.08\times10^{12}$. Only the Cs atoms excited by the 455 nm pumping laser can be involved in 1359 nm stimulated emission output. According to our previous work~\cite{Xu2}, the 455 nm pumping laser beam is aligned parallel to the main cavity mode, and therefore a velocity spread of 10.74 m/s parallel to the main cavity mode of Cs atoms can be excited to the 7S$_{1/2}$ state. The velocity distribution function of Cs atoms in the main cavity mode is $f(v_x)=\sqrt{\frac{m}{2\pi k_B T}}\exp (\frac{-m{v_x}^2}{2k_B T})$, where $m=2.22\times10^{-25}$ kg is the mass of Cs atoms, while $v_x$ is the velocity component parallel to the main cavity mode of Cs atoms. The effective number of Cs atoms in the main cavity mode that can be excited to the 7S$_{1/2}$ state, i.e. the number of Cs atoms can be involved in the 1359 nm stimulated emission output of 633 nm and 1359 nm good-bad cavity dual-wavelength active optical clock, is thus calculated to be $N_{eff}=Nf(v_x=0)\Delta v_x=1.08\times10^{11}$, where $\Delta v_x$ is the 10.74 m/s velocity spread.

The atomic interaction time with the main cavity mode $t_{int}=\frac{1}{\gamma_{34}+\gamma_{35}+\gamma_{41}}=21.6$ ns, where $\gamma_{34}=6.23\times 10^6$ s$^{-1}$, $\gamma_{35}=11.4\times 10^6$ s$^{-1}$, $\gamma_{41}=28.6\times 10^6$ s$^{-1}$, and $\gamma_{ij}$ represents the corresponding spontaneous emission rate between the state ~$ |i\rangle $ and ~$ |j\rangle $ as shown in Fig.1. The cycle time of a Cs atom involved in the 1359 nm stimulated emission is $t_{cyc}=\frac{1}{\Omega}+\frac{1}{\gamma_{23}}+\frac{\pi}{2g\sqrt{n+1}}+\frac{1}{\gamma_{41}}=300$ ns$+\frac{\pi}{2g\sqrt{n+1}}$, where $\gamma_{23}=4.05\times10^6$ s$^{-1}$, and n is the average photon number inside the cavity. The photon number rate equation is~\cite{Xue}£¬
\begin{equation}
\frac{dn}{dt}=N_{eff}\frac{\rho_{33}-\rho_{44}}{t_{cyc}}sin^2(\sqrt{n+1}gt_{int})
-\kappa n\end{equation}
Where $\rho_{33}=5.8\%$ and $\rho_{44}=1.4\%$~\cite{Yanfei} are population probability of 7S$_{1/2}$ and 6P$_{1/2}$ state respectively. In the steady state, the average photon number is $n=2.84\times10^6$ from Eq.1. The quantum limited linewidth of the 1359 nm output is~\cite{Kuppens-1,Yu2,Chen2}
\begin{equation}
\Delta\nu=\frac{\Gamma_{c1359}}{2n}\frac{\rho_{33}}{\rho_{33}-\rho_{44}}(\frac{1}{1+a})^2
\end{equation}
where $\Gamma_{c1359}= 217.39$ MHz is the 1359 nm main cavity mode linewidth. The 1359 nm quantum limited linewidth of good-bad cavity dual-wavelength active optical clock is thus calculated to be 77.6 mHz.

So far, a linewidth of 250 mHz has been realized with PDH technique near room temperature~\cite{Jiang}. If a linewidth of 250 mHz was realized in 633 nm PDH frequency stabilization  system of good-bad cavity dual-wavelength active optical clock, the linewidth of main cavity-length noise at 1359 nm is then 116 mHz. The 1359 nm output linewidth can thus be reduced to 4.55 mHz due to the suppressed cavity-pulling effect~\cite{Xu2}, far narrower than the 77.6 mHz quantum limited linewidth of good-bad cavity dual-wavelength active optical clock. Therefore, it is feasible to experimentally realize the 77.6 mHz quantum limited linewidth. All the theoretical discussion here does not include the influence of collision frequency shift. The relevant experimental data of Cs 1359 nm transition is still unavailable and further research is needed to estimate the influence of collision frequency shift on 1359 nm output linewidth.

\section{Conclusion}
In this paper the dual-wavelength active optical clock is experimentally demonstrated for the first time. The output power of Cs four-level active optical clock is measured when changing the Cs cell temperature. Both 1359 nm and 1470 nm stimulated emission output of Cs four-level active optical clock are detected between the 118 $^{\circ }C$ and 144 $^{\circ }C$ Cs cell temperature. A second experimental setup of Cs four-level active optical clock pumped by a 459 nm laser is established and the 1470 nm stimulated emission output linewidth The 1470 nm stimulated emission output linewidth of each experimental setup of Cs four-level active optical clock is 590 Hz is measured to be 590 Hz by beating these two independent setups pumped by 455 nm and 459 nm lasers separately. The main cavity length is not stabilized and the frequency stability of 1470 nm stimulated emission output is greatly affected by the perturbation from external environment. To lock the main cavity length of active optical clock, we propose the experimental scheme of 633 nm and 1359 nm good-bad cavity dual-wavelength active optical clock. The 633 nm stimulated emission is designed to work in the good-cavity limit, while the 1359 nm stimulated emission is working in the bad-cavity regime. The 633 nm output frequency is locked to a super-cavity with the Pound-Drever-Hall (PDH) technique ~\cite{Drever, Young, Holger, Jiang, Kessler} and the cavity length is thus stabilized. Due to the suppressed cavity pulling effect in active optical clock, the frequency stability of 1359 nm bad-cavity stimulated emission output is then expected to be further improved by at least 1 order of magnitude than the 633 nm PDH system. The cavity pulling effect is designed to be reduced by a factor of about 25.5 for 633 nm and 1359 nm good-bad cavity dual-wavelength active optical clock and the quantum limited linewidth of 1359 nm output is estimated to be 77.6 mHz.

This research was supported by the National Natural Science Foundation of China (Grant Nos. 10874009 and 11074011), and International Science $\&$ Technology Cooperation Program of China under No. 2010DFR10900.

\end{document}